\documentstyle[aps,prl,epsf,floats,twocolumn]{revtex}
\begin{document}
%\draft

%\twocolumn
\wideabs{
\title{Contributions of spontaneous phase slippage
to linear and non--linear conduction near the Peierls
transition in thin samples of {\it o} -TaS$_3$}

\author{V.~Ya.~Pokrovskii and S.~V.~Zaitsev--Zotov\\}
\address{Institute of
Radioengineering and Electronics of Russian Academy of
Sciences, Mokhovaya 11,
103907 Moscow,
Russia, e-mail: pok@mail.cplire.ru\\}
%\date{\today}
\maketitle

\begin{abstract}
In the Peierls state very thin samples of TaS$_3$
(cross-section area  $\sim 10^{-3}$~$\mu$m$^2$) are found to demonstrate
smearing of the I-V curves near
the threshold field.
With approaching
the Peierls transition temperature, $T_P$, the smearing 
evolves into
smooth growth of conductance from zero voltage
interpreted by us as the contribution of fluctuations to the non--linear 
conductance. We
identify independently the
fluctuation contribution to the linear conductance near
$T_P$. Both linear and non-linear contributions depend
on temperature with 
close activation energies $\sim (2$\ ---\ $4) \times 10^3$~K and apparently reveal 
the same process. We
reject 
creep of the {\it continuous} charge-density waves (CDWs)
as the origin of this effect and show that it is spontaneous phase slippage 
that results in creep of the CDW. A model is proposed
accounting for both the linear and non-linear parts of the fluctuation 
conduction
up to $T_P$.
\end{abstract}
} %wideabs

%\twocolumn
%\samepage

%\narrowtext

\section{Introduction}
Though the Peierls
transition in quasi 1-dimensional conductors was discovered more than 30 years 
ago, its mechanism and the role of fluctuations still
remain an unsettled problem \cite{grobzor,mobzor}. 
The fluctuations in quasi 1-dimensional compounds below $T_P$ are seen from 
various studies
including transport properties, such as the large width of the Peierls transition 
\cite{grobzor,mobzor} in comparison with that expected from the BCS-type onset 
of the gap, the smeared edge of the Peierls gap revealed through the optical 
investigations \cite{ItNa}, the appearance of spontaneous 
current noise, which is associated
with thermally initiated phase slip (PS) developing several kelvins below
$T_P$ \cite{fluctran}.

A decrease of the cross-section area $s$ of the samples results in growth of the 
fluctuations. For example, in the samples of o-TaS$_3$ with
$s \sim 10^{-2}$~$\mu$m$^2$ and below, the Peierls transition is 
smeared out and substantially shifted down to lower 
temperatures \cite{borzzn}. Conductance 
hysteresis in such thin samples is absent within
decades of kelvins below $T_P$; in this temperature range spontaneous
PS is observed, and the conductivity strongly
deviates from the Arrhenius law \cite{fluctran,pisfluct}.
 Another fluctuation effect known as threshold rounding consists in
smearing out of the onset of the non-linear current at
the threshold field $E_T$ \cite{Gilround,thornround1,thornround}. 
This effect is found in NbSe$_3$. The rounding 
increases both with increasing $T$ and decreasing thickness
$t$ of the crystals; in the thinnest crystals the
growth of conductivity starts from zero
field. In \cite{Gilround,thornround1,thornround} phase slippage has been
discussed as a possible basis of the rounding, but
the authors did not find enough arguments in favor of this explanation. Another
interpretation was found
to be more reasonable \cite{Gilround,thornround1,thornround}: 
the rounding was attributed to the thermally--assisted creep of charge--density 
waves (CDWs) in the
framework of the weak-pinning model \cite{grobzor,mobzor}. 
This approach implies that
in very thin samples the pinning energy of the phase-correlation volume
becomes comparable with $kT$, 
and activated creep of the continuous CDW within 
the correlation lengths $L_{2\pi}$
is possible. Estimates for NbSe$_3$ based on the mean-field
BCS dependence for $\Delta(T)$, showed that thermal depinning of the CDW 
is probable. This interpretation, however, is rather dubious for TaS$_3$, where 
the mean-field approach fails near $T_P$: in highly anisotropic compounds such 
as TaS$_3$ and K$_{0.3}$MoO$_3$, the onset of the gap near $T_P$ does not follow 
the BCS dependence \cite{xray,lera}, and the pseudogap does
not vanish tens of kelvins above $T_P$ \cite{ItNa,gorshunov,zzztex}.

In the present paper the
threshold rounding in thin samples of TaS$_3$ is reported.
Independently we observe a fluctuation contribution to the linear conductivity.
It is shown that
creep of the {\it continuous} CDW
cannot account for the threshold rounding in
TaS$_3$. Alternatively, we show
that spontaneous PS observed near $T_P$ results in
local creep of the CDW and contributes to the linear and non--linear 
conductivity, in agreement with our
experiment. The result is generalized for large
samples. We discuss the mechanism of the Peierls
transition in the light of the PS --
induced creep.

\section{Experimental Technique and Results}
Thin samples of TaS$_3$ were placed on sapphire
substrates. We used vacuum--deposited indium contacts 
\cite{borzzn86}. The cross--section area of the samples
was estimated from the values of the room-
temperature resistance ($3\times 10^{-4}$~$\Omega
$cm) and the visible contact separation \cite{borzzn86}. Similar results are
observed on 5 samples from high-quality
batches. Most of the data reported here are obtained on the representative 
sample with the dimensions
$L=4.5$~$\mu$m, $s=0.3 \times 10^{-3}$~$\mu$m$^2$.

\begin{figure}
\vskip -5cm
\epsfxsize=11cm
\centerline{\epsffile{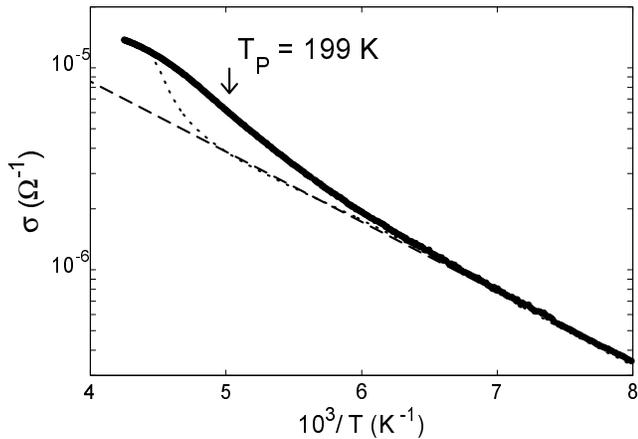}}
\vskip -4.5cm
\caption{Temperature dependence of the low-field resistance for the 
representative sample measured at $V=20$~mV ($T<178$~K) and at $V=10$~mV 
($T>178$~K). The broken line, $\sigma_\Delta$, shows the extrapolation 
of the low-temperature conductivity $\sigma_\Delta\propto \exp(\Delta/T)$, 
$\Delta =800$~K. The dotted line is normalized to a low temperature 
$\sigma(T)$ curve for a typical thick pure sample.}
\label{sigmat}
\end{figure}

\begin{figure}
\vskip -4.6cm
\epsfxsize=13cm
%\centerline{
\epsffile{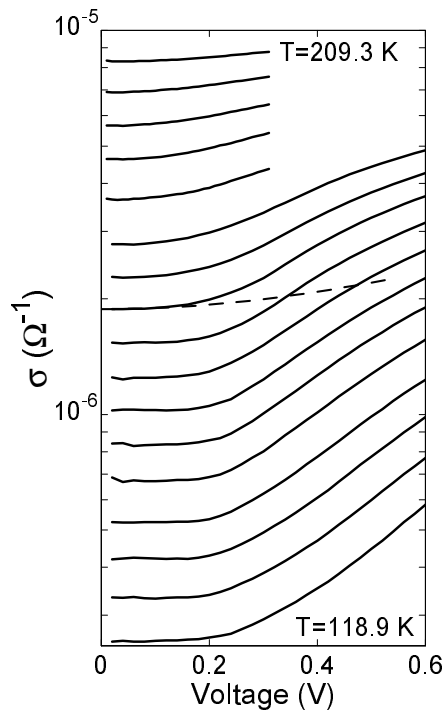}
%}
\vskip -4.5cm
\caption{Voltage dependencies of the conductivity ($I/V$) at 
$T=$~ 209.3, 203.4, 197.4, 191.4, 184.5, 176.8, 171.2, 165.9, 160.1, 154.6, 
149.3, 144.3, 
139.1, 133.6, 128.8, 123.9, 118.9~K. The broken line shows an example of the fit 
of $\sigma_{nl}(E)$ by Eq.~(\protect\ref{lnl}), where $E_T=480$~V/cm, $\sigma_l = 
\sigma-\sigma_{\Delta}$ (see text).}
\label{sigmav}
\end{figure}

\begin{figure}
\vskip -3.5cm
\epsfxsize=10cm
\centerline{\epsffile{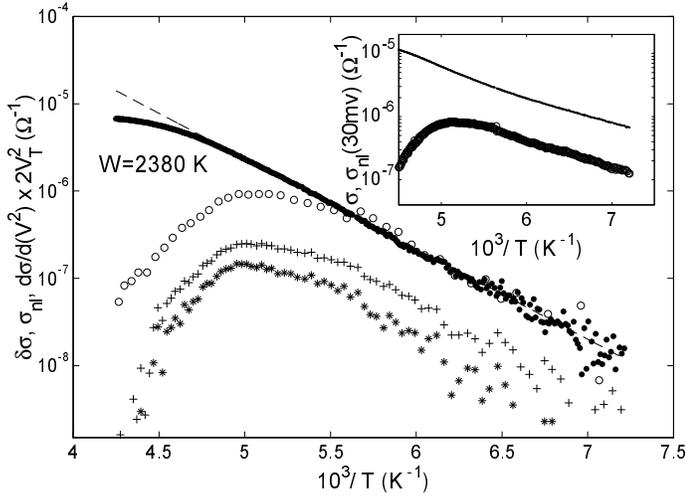}}
\vskip -2.5cm
\caption{Temperature dependencies of the excess linear conductivity (dots), non-
linear conductivity for two values of $V$ below $V_T$:~'$\ast$' - $V=120$~mV, 
'+' - $V=160$~mV. The open circles show the fit of the non--linear conductivity 
in accordance with Eq.~(\protect\ref{lnl}) with $V_T=220$~mV ($E_T=480$~V/cm)
 (see text). The inset shows the total conductivity (dots) together with 
non--linear conductivity (circles) at fixed $V=300$~mV ($V>V_T$).}
\label{sigmaflt}
\end{figure}

The dependencies of conduction $\sigma$ on temperature and voltage $V$ are 
presented in
Figs.~\ref{sigmat} and \ref{sigmav}
respectively. One can see (Fig.~\ref{sigmat})
that the Peierls transition is 
smeared out in comparison with the transitions in
usual-sized samples (shown with a dotted line),
in agreement with Ref.~\cite{borzzn}. Deviation
from the Arrhenius law is observed tens of
kelvins below $T_P$ (indicated by an arrow), the
latter being considerably shifted downwards \cite{borzzn} in
comparison with $T_P=220$~K observed in thick
samples. The activation dependence 
$\sigma \propto \exp(-\Delta/T)$ with $\Delta = 800$~K
extrapolated from the low temperatures is shown
by the broken line; we denote the corresponding
conductivity as $\sigma _{\Delta}$. 
We shall consider the temperature and sample-size dependence of the Peierls gap, 
$2\Delta$, to be insignificant, which is supported by the results of
Refs.~\cite{ItNa,thornround,zzztex}. Then, the difference
$\delta\sigma \equiv 
\sigma - \sigma _{\Delta}$
can be considered as the fluctuation contribution
to the conductivity. The Arrhenius
plot $\delta\sigma$ {\it vs.} $1/T$ is shown in
Fig.~\ref{sigmaflt}. The
dependence is close to a straight line up to $T
\approx T_P$; the activation
energy $W$ being about 2400~K, is well above $\Delta$. 
For samples with higher cross--section 
areas we obtained somewhat larger activation energies,
up to $W \approx (5-7)\times10^3$~K for the normal--sized samples,
as it was reported earlier \cite{leningr}.

Turning to the $\sigma$ {\it vs.} $V$ dependences
(Fig.~\ref{sigmav}) we note that the onset of the non--linear
 conduction is smeared,
the threshold rounding being
clearly seen above $T=120$~K. The scale of the 
voltages applied corresponds to large fields, above 1~kV/cm.
At lower temperatures (about 120~K) the onset of the non--linear
conduction is relatively sharp, and we can estimate the 
threshold for collective conduction as $V_T \approx 0.2$~V 
($E_T \approx 400$~V/cm), in accordance with the size
effect \cite{borzzn}. We shall assume that at higher temperatures
the value of $E_T$ is approximately the same. Thus, the non--linear conduction
for $V < 0.2$~V will be referred to as subthreshold non--linear conduction, while
at $V$ well above 0.2~V collective conduction is expected. 
Below we shall see that this division is not unphysical.

One can see (Fig.~\ref{sigmav}) that the rounding progresses
with approaching $T_P$. At high temperatures ($T \gtrsim 170$~K) it is 
impossible to
 define a voltage range of linear
conduction: the non--linearity starts from zero
voltage. Fig.~\ref{sigmaflt} shows the non--linear
conductance, $\sigma_{nl} \equiv \sigma (V)-
\sigma(0)$, at fixed values $V<V_T$ as a function
of $T$, together with $\delta \sigma(T)$.
Evidently, $\delta \sigma(T)$ and $\sigma_{nl}(T)$ behave in 
similar ways up to $T \approx 175$~K, while at higher
temperatures $\sigma_{nl}$ deviates downwards. One can conclude that the excess
conductivity, $\delta\sigma$, and the threshold
rounding have a common underlying mechanism at least at the lower temperatures.
The possibility of coupling of the
non-linear conduction below $E_T$ with an enhancement of low-field
conductivity was also noticed in Ref.~\cite{Gilround} for thin samples of NbSe$_3$.

The scaling
observed resembles that between the linear and
non--linear conduction {\it above} $V_T$ \cite{scaling} known 
for different CDW conductors,
including TaS$_3$ \cite{tascaling}. Our samples also
demonstrate such a scaling: the inset to
Fig.~\ref{sigmaflt} shows the dependence
$\sigma(T)$ together with $\sigma_{nl}(T)$ at
fixed $V>V_T$. In agreement with earlier observations, both values depend on 
temperature in a similar way, with the activation energy $\sim \Delta$. It is 
clear that the scaling between $\sigma_{nl}(T)$ ($V<V_T$) and $\delta \sigma(T)$ 
is
quite different, as the slopes of the curves
correspond to much higher activation energies,
$W \gg \Delta$. 

\section{Discussion}
The large values of $E_T$ result from the small 
transversal dimensions of the samples, in accordance with 
the size effect 
observed in TaS$_3$ \cite{borzzn}. As both transversal dimensions
of our TaS$_3$ samples are of the same order of
magnitude \cite{borzzn}, we expect the pinning to be 1-dimensional,
rather than 2-dimensional, as in the thin samples of NbSe$_3$ 
\cite{Gilround,thornround,reject}.
Following the explanations of rounding in NbSe$_3$ 
\cite{Gilround,thornround1,thornround}, we could assume that the pinning 
energies of
phase--correlation volumes in TaS$_3$ nanosamples
are small enough to enable thermal depinning of
phase--coherent volumes. 

The lowest-energy local depinning of the CDW results in a phase gain
by the value 
$\sim 2\pi$ over the phase-correlation length
\cite{Littlewood}. Such a deformation will cause a
CDW stress, resulting in a local variation of
resistance by percents for our samples \cite{critstat}.
Meanwhile, metastable states cannot
exist in such thin samples in the vicinity of $T_P$:
the hysteresis loop develops only below 140~K
in the representative sample (in usual-sized pure samples the hysteresis 
develops 5---7~K below $T_P$ \cite{fluctran,Higill}). 
So, any deformation immediately relaxes via a PS act, {\it i.e.}
a plastic deformation of the CDW.
The PS act is followed, sequentially, by local creep of the CDW \cite{critstat}. 
This results in a phase perturbation of the
same order of magnitude as the initial elementary act of
creep \cite{critstat}, and so PS is to be taken into
account.
Below we discuss in detail the
conditions for the spontaneous PS \cite{fluctran,pisfluct,epl} and its effects 
on the
conductivity.

Remarkable is that the activation energy for the fluctuations is nearly 
independent of the field applied while it is below $E_T$: the slopes of the 
excess linear conductivity and of the non--linear conductivity at $V=160$~mV (which is 
quite close to $V_T \sim 200$~mV) are close, while the activation energy for the 
creep should run to zero at $E \to E_T$ \cite{Natter}. So the process initiating 
the fluctuation conduction is other than creep of the CDW. At the same time, at 
low temperatures ($\lesssim 130$~K) the non--linear conductivity
becomes distinguishable only close to $E_T$, {\it i.e.}
reveals itself as the threshold rounding. So $E_T$ is a characteristic field 
for the fluctuation conductivity, and the latter is in a way
coupled to the CDW creep. This apparent contradiction is removed by the the 
following consideration.

Evidently, the mechanism initiating the conductivity is the PS: the high 
activation energy is typical for PS in TaS$_3$ \cite{fluctran,borzzn}, and its 
independence of $E$ at $T>120$~K was reported in Ref.~\cite{bottle}. At the same time 
according to Ref.~\cite{critstat} each PS act is followed by temporary creep 
(rearrangement) of the CDW in the vicinity of the point where the PS occurred.
In the presence of an external electric field the creep prevails in the 
direction defined by the field and provides a mechanism of the CDW conduction 
below $E_T$. At $E \to E_T$ the CDW phase-correlation length diverges 
\cite{Wagong}, so $E_T$ is expected to be the critical point for the conduction. 
A hypothesis that phase slippage (in particular, edge dislocations) could 
facilitate CDW creep was also remarked in Ref.~\cite{Gilround}.

\begin{figure}
\vskip -4.2cm
\epsfxsize=10cm
\centerline{\epsffile{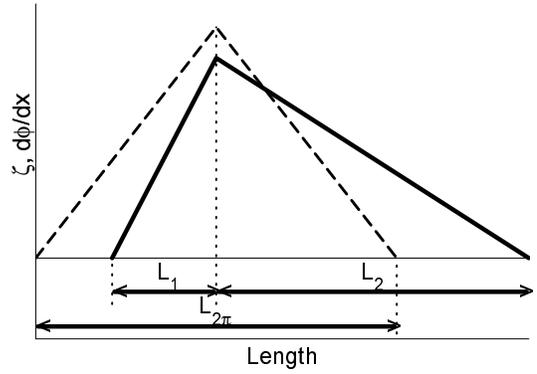}}
\vskip -4.5cm
\caption{Phase ($\Phi$) redistribution following a PS act at zero (broken line) 
and non--zero field $E=\frac{1}{2}E_T$ (solid line) according to the critical-state model. 
The PS-induced creep results in an
average shift of the charge density to the right.}
\label{procol}
\end{figure}

In the case of 1-dimentional pinning (which could be applied to our samples) and 
PS involving the whole cross-section area, we can estimate the current induced 
by 
the PS. For simplicity let us consider the initial state to be uniform, {\it 
i.e.} the shift of the chemical potential $\zeta=$const. Entering of a new 
period in the absence of external field is followed by CDW
creep under the internal electric fields
$E_{int}=d\zeta/dx$. The creep proceeds while
the effect of $E_{int}$ exceeds the effect of impurities, which we 
for simplicity describe by the average value, $E_T$. The resulting phase 
perturbation
(Fig.~\ref{procol}) covers the length  \cite{critstat}
\begin{equation}
\L_{2\pi} \approx 2\sqrt{\pi(d\zeta/dq)/E_T},
\label{L2pi}
\end{equation}
where $d\zeta/dq$ characterizes the CDW elastic
modulus, $q$ being the in-chain component of the CDW wave vector. Note, that 
$L_{2\pi}$ appears to be
of the order of the phase-correlation length \cite{grobzor,critstat}.
 Under an external electric
field $E<E_T$ the creep proceeds
asymmetrically with respect to the point of the PS
nucleation giving the
divergence of $L_{2\pi}$ at $E \to E_T$ \cite{critstat}.
The new period is
distributed so that the $d\zeta/dx=E_T+E$ from one side of the maximum remnant 
deformation and
$-(E_T-E)$ from the other side (Fig.~\ref{procol}).
The resulting progress of the CDW 
(and of the coupled charge $2e$ per chain) in the direction defined by $E$ could 
be estimated as $\delta L
= \frac{1}{3}(L_2-L_1)$, where $L_1$ and $L_2$ are the lengths of the phase 
perturbations in the two directions (Fig.~\ref{procol}) \cite{procol}. 
With the condition that the areas under the triangles 
(Fig.~\ref{procol}) should be equal and correspond with the phase gain $2 \pi$ 
we obtain from  simple calculations:
\begin{equation}
\delta L(E) = \frac{1}{3}L_{2\pi}\frac{E}{E_T}\frac{1}{\sqrt{1-(E/E_T)^2}}.
\label{dL}
\end{equation}
If the PS nucleation rate per unit length is 
$\nu(T,E)$, then the resulting mean current is
\begin{equation}
I_{PS}=2e\nu \delta L
\label{I}
\end{equation}
per chain. As each PS act (fluctuator) affects the length $\sim L_{2 \pi}$,
$L_{2 \pi} \nu \equiv f$ may be considered as a typical frequency
of switching of independent fluctuators. The temperature dependence of the PS 
rate could be empirically presented as
$\exp(-W/T)$ \cite{rama}, where $W \sim
(5$\ -\ $7) \times 10^3$~K \cite{fluctran,borzzn,gillPS}.
So, Eqs.~\ref{dL} and \ref{I}
give the dependence of the excess current both on $T$
and $E$. As $E \to E_T$ an unphysical
divergence of $I_{PS}$ occurs, because in the model
we have neglected the time of creep, $\tau_{cr}$, following each PS act.
% in comparison with the reverse frequency of the PS. 
With approaching
$E_T$ $\tau_{cr}$ grows together with $L_2$ (Fig.~\ref{procol}), and the PS 
frequency becomes dominated by $1/\tau_{cr}$. At low temperatures when $f$ is 
relatively  small, 
$I_{PS}$ becomes noticeable only for $E$ close to $E_T$:
Eqs.~\ref{dL} and \ref{I} thus feature the threshold rounding. At higher $T$ (and 
$f$) the
area of validity of Eq.~(\ref{I}) shrinks down to smaller
$|E|$. In the limit of small
$|E|$ neglecting the dependence of $\nu$ on $E$ \cite{bottle} we obtain:
\begin{equation}
I_{PS}=\frac{2}{3}ef \left[ \frac{E}{E_T}+\frac{1}{2}
\left( \frac{E}{E_T} \right) ^3 \right] \equiv I_l + I_{nl}.
\label{Ipar}
\end{equation}
Thus, spontaneous PS gives contributions both to linear
($I_l$) and non--linear ($I_{nl}$)
currents. 
Note that extrapolation of $E$ to $E_T$ gives
a relation between $I_{PS}$ and $f$ that is very similar to
that between the CDW current and the fundamental
frequency. This is natural, because for $E \to E_T$ each
pair of electrons entering  the CDW via a PS act creeps along
the whole sample.  

From
Eq.~(\ref{Ipar}) we obtain 
\begin{equation}
\sigma_{nl}=\frac{1}{2}\sigma_l\frac{E^2}{E_T^2},
\label{lnl}
\end{equation}
where $\sigma_l $ is the PS--induced part of the linear conductivity. At fixed 
$E/E_T$ ($E<E_T$)
the fluctuation non--linear conductivity should have the same
temperature dependence as the linear conductivity: both are
dominated by $f \propto \exp(-W/T)$ [Eq.~(\ref{Ipar})]. Neglecting
the dependence of $E_T$ on $T$  we come to the scaling between
$\delta\sigma$ and $\sigma_{nl}$, as the one observed from
Fig.~\ref{sigmaflt}. For a quantitative comparison of $\delta\sigma$ and 
$\sigma_{nl}$ note that $[d\sigma/d(E^2)] \times 2E_T^2$ is simply 
$\sigma_l$~($\equiv \delta \sigma$) [Eq.~(\ref{lnl})]. 
To check this, 
we show the value $[d\sigma/d(E^2)] \times 2E_T^2$ in Fig.~\ref{sigmaflt},
where $d\sigma/d(E^2)$ is determined from the best fit of 
$\sigma(V)$ ($V\ll V_T$) with parabolic
dependencies and $E_T$ is a fitting parameter. An example of the fit of 
$\sigma(V)$ by Eq.~(\ref{lnl}) is shown in Fig.~\ref{sigmav} with a broken line. 
For the representative sample we get $E_T=480$~V/cm.
For different samples, the values obtained from the fit by Eq.~(\ref{lnl}) agree 
with the results of direct measurements, though they are
somewhat larger \cite{hiEt}. 

\begin{figure}
\vskip -4.1cm
\epsfxsize=10cm
\centerline{\epsffile{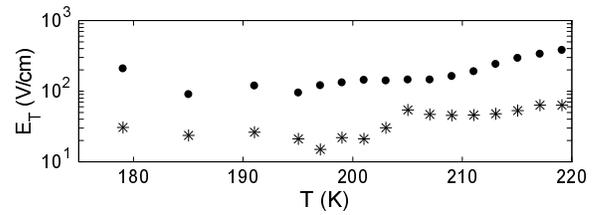}}
\vskip -7cm
\caption{The $T$ dependence of $E_T$ for the sample with 
$L=4$~$\mu$m, $s=1.5 \times 10^{-3}$~$\mu$m$^2$. $\ast$ --- direct measurements, 
black circles
--- from the values of $d\sigma/d(E^2)$ and $\delta \sigma$ (see text).}
\label{EtEt}
\end{figure}

Additional evidence of the correlation between the threshold field and 
non--linear 
fluctuation conduction is provided by the measurements of another sample 
with approximately the same length but larger cross-section area, $s=1.5 \times 
10^{-3}$~$\mu$m$^2$, and somewhat larger activation energy for $\delta 
\sigma(T)$,
$W=3400$~K. A similar treatment of $\sigma_{nl}$ with the help of Eq.~(\ref{lnl})
gave us the dependence $E_T(T)$. In addition, we were able to measure $E_T(T)$ up to 
$T=T_P$ and higher \cite{zzztex}
directly, as the onset of sharp non--linear conduction.  This was possible after 
subtracting the part of conductivity
$\propto E^2$ from each $\sigma(V)$ curve \cite{zzztex}. The values of $E_T$ 
determined both ways
are presented in Fig.~\ref{EtEt} as a function of temperature. Both dependences 
show mesoscopic-type irregular variations 
of $E_T$ with temperature \cite{zzpokgill}, though $E_T$ obtained from 
Eq.~(\ref{lnl}) is about 5 times larger \cite{hiEt}. One can see correlation 
between the two dependences.

From Eq.~(\ref{Ipar}) we obtain reasonable estimates of the 
frequencies of switching for the fluctuators. To observe distinct excess 
conductivity (at 140~K for the representative sample,
Fig.~\ref{sigmaflt}), we should take $f\sim 10^7$~Hz.
This
is only 2---3 orders of magnitude higher than we were able to 
see directly from the time domains of the fluctuations \cite{pisfluct,epl}, the 
latter 
being restricted 
by the electric scheme. Thus, both the values and the activation
energy for the linear and non--linear fluctuation currents
are fairly described by the model proposed.

It is worth
mentioning that the dependence of $|d^3V/dI^3|$ {\it vs} $T$
below $T_P$ presented in Ref.~\cite{Gilround} fits the Arrhenius
law fairly well with $W \approx 4500$~K, in agreement with the 
PS measurements at the contacts \cite{gillPS}.

The lowering of $W$ 
with the decrease of the sample's
cross-section area also finds natural explanation within the model proposed. In 
fact, a large threshold field corresponds to high inhomogeneous stress of the 
CDW in the thin samples \cite{critstat}:
\begin{equation}
<\zeta^2>^{1/2} \sim 2\sqrt{\pi E_T (d\zeta/dq)},
\label{zeta}
\end{equation}
The stress lowers the barrier for the PS in certain points \cite{epl}. The 
decrease of the sample cross-section area reveals itself in the growth of $E_T$, 
and thus results in the lowering of the activation energies for $\delta 
\sigma(T)$ and $\sigma_{nl}(T)$. Earlier
we have explained in a similar way the broadening of the range of the
fluctuations and of the Peierls transition along the temperature
scale in the thin samples \cite{epl}. Note that the model of thermal depinning
of the phase-coherent volumes \cite{Gilround,thornround} gives a much stronger 
size
dependence of the excess conductivity: the
depinning energy is proportional to $s^{2/3}$ (Ref.~\cite{thornround}). So, for the 
sample with 
$s=1.5 \times 10^{-3}$~$\mu$m$^2$ we should expect 
$W$ to be about 7000~K (instead of 3400~K), and 
the excess conductivity should become negligible in the
thick samples. Actually, we found no qualitative
difference  between the excess conductivity of the thick and thin samples. The 
activation
energy for $\delta \sigma(T)$ in thick samples is 
$(5-7)\times 10^3$~K \cite{fluctran,leningr}, in agreement
with the activation energy found from the noise probing of the spontaneous PS 
\cite{fluctran}.
The threshold rounding is also noticeable
in thick samples \cite{mobzor,zzztex}, though the study of the  
non--linear fluctuation conductivity
is complicated because of its narrow temperature range
and small $E_T$.

Note that the dependence $\delta \sigma(T)$ follows the 
activation law up to $T_P$, and even a little bit 
above it (Fig.~\ref{sigmaflt}),
no feature being observed at $T_P$. So,
the state a little bit above $T_P$ could be considered as a CDW saturated with 
climbing dislocations rather than a single--electron
state. The conduction of such a mixture is supplied by
the processes of nucleation and motion of the domain 
boundaries, which exert high internal electric fields to
the domains. The fact that the dependencies characterizing the non--linear 
conductivity deviate from the Arrhenius law at lower temperatures than $\delta 
\sigma(T)$ could be ascribed to the growth of $E_T$ near $T_P$ 
(Fig.~\ref{EtEt}); note also that with increasing $T$ the model fails first 
at finite $E$, and then at $E \to 0$.

In conclusion, we have observed the fluctuation contribution to the conductivity 
of thin samples of {\it o}-TaS$_3$, which comprises linear and non--linear parts. 
We have shown that the spontaneous phase slippage observed in the CDW in the 
vicinity of $T_P$ results to the excess conductivity whose temperature and 
electric-field dependences match our experimental observations. 

The simple model proposed requires further development. In particular, the 
mechanism of PS nucleation and evolution should be considered in terms of 
nucleation and propagation of dislocation loops in the CDW \cite{Rama}.
%; the internal stress field of the CDW should be taken into account. 
A possible contribution of glide of the dislocation lines to the current 
\cite{gillepl,ertecrys} also requires analysis. In the case of bulk (3D) 
samples the loops cover only part of the cross-section areas, so 
transversal interaction of the chains while the local creep proceeds should be 
considered. A special case is the ribbon--like (2D) samples 
evidently treated in 
Refs.~\cite{Gilround,thornround1,thornround}. A dislocation loop degenerates into a 
pair of points interacting logarithmically. Then the system acquires the 
features of a 2D crystal exhibiting the Kosterlitz-Thouless transition 
\cite{kost}. This approach can explain the lowering of $T_P$ in thin crystals and 
gradual power--law dependences of $\sigma_{nl}$ on $(T)$.

\section{Acknowledgments}

We are grateful to P. Monceau for help in the experiment, 
to Yu.~I.~Latyshev, Ya.~S.~Savitskaya,
and V.~V.~Frolov for
producing the samples,
and to S.N.~Artemenko and A.A. Sinchenko for fruitful
discussions.
This work was supported by the
Russian Foundation for Basic Research 
(grants Nos. 98-02-16667, 99-02-17387), Jumelages 19 
(RFBR, grant No. 98-02-22061), and 
MNTP ``Physics of Solid State Nanostructures'' (grant No. 97-1052).


\begin{references}
\bibitem{reject}Initially alternative explanations of the effect of cross-
section area on $E_T$ have been proposed (Refs.~\cite{borzzn,gill}). In 
Ref.~\cite{thornround} 
one can find a detailed discussion rejecting these possibilities.
\bibitem{procol}We note the simplifications of the model: we have neglected 
the inhomogeneity of $\zeta (x)$ preceding the PS act, while actually a PS 
nucleates at the point of the highest CDW stress (maximum of $|\zeta|$). Also, 
there might be a shift of the maximum of 
the remnant deformation (the summit of the triangle in 
Fig.~\protect\ref{procol}) with 
respect to the point of the PS initiation. So the factor 1/3 could be not 
exact.
\bibitem{rama}We do not reject the model Ref.~\cite{Rama}, according to which the PS 
results from the nucleation of a dislocation loop of critical radius. However, 
this model does not give spontaneous PS in large pure samples and needs 
additional specifications to describe the loops in real samples. The simple 
one--barrier model looks good enough to describe empirically the PS dependence on $T$ 
without deepening into its microscopics.
\bibitem{hiEt}In the model above we approximated the interaction of the CDW with 
the impurities by an average value $E_T$. Actually, $E_T$ forms over the 
phase--correlation length scale, which could exceed $L_{2\pi}$ 
[Eq.~(\ref{L2pi})] by a 
unity factor, while we
apply the approximation over the length $ L_{2\pi}/2$.
On this length scale a higher value of $E_T$ is expected, in accordance with the 
size effect in 0D pinning \cite{zzpokgill}.
\end{references}
\end{document}